# IMAGESPACE: AN ENVIRONMENT FOR IMAGE ONTOLOGY MANAGEMENT

Shiyong Lu, Rong Huang, Artem Chebotko, Yu Deng, and Farshad Fotouhi

*Abstract*: *More and more researchers have realized that ontologies will play a critical role in the development of the Semantic Web, the next generation Web in which content is not only consumable by humans, but also by software agents. The development of tools to support ontology management including creation, visualization, annotation, database storage, and retrieval is thus extremely important. We have developed ImageSpace, an image ontology creation and annotation tool that features (1) full support for the standard web ontology language DAML+OIL; (2) image ontology creation, visualization, image annotation and display in one integrated framework; (3) ontology consistency assurance; and (4) storing ontologies and annotations in relational databases. It is expected that the availability of such a tool will greatly facilitate the creation of image repositories as islands of the Semantic Web.*

## 1. Introduction

More and more researchers have realized that ontologies will play a critical role in the development of the Semantic Web, the next generation Web in which content is not only consumable by humans, but also by software agents. [1,5]. Undoubtedly, images will be major constituents of the Semantic Web, and how to share, search and retrieve images on the Semantic Web is an important but challenging research problem. Unlike other resources, the semantics of an image is implicit in the content of an image. Although this is not a problem to human cognition, it imposes a challenge on image searching and retrieval based on the semantics of image content. Manual annotation of images provides an opportunity to make the semantics of an image explicit and richer. However, different annotators might use different vocabulary to annotate images, which cause low recall and precision in image search and retrieval. We propose an ontology-based annotation approach, in which an ontology is created for a particular domain so that the terms and their relationships are formally defined. In this way, annotators of a particular image domain, say, the Family Album domain, will use the same vocabulary to annotate images, and users will search images guided by the ontology with greater recall and precision.

In [11, 12], we have briefly described *ImageSpace*, an image ontology creation and annotation tool, and our experience of annotating linguistic data using *ImageSpace* for the preservation of endangered languages [13, 17, 18]. This paper extends these results with ontology visualization, the storage of ontologies and annotations in relational databases, and ontology-based information retrieval. In summary, the contributions of this paper are:

- *ImageSpace* supports the functionality of ontology creation. In particular, it facilitates the creation of classes, properties, and relations between classes and relations between properties. It also provides ontology consistency assurance;
- *ImageSpace* provides full support for the standard web ontology language DAML+OIL [2];
- *ImageSpace* supports the visualization of an ontology to enable users to navigate, zoom-in and zoom-out various portions of an ontology.
- *ImageSpace* supports ontology-driven annotation of images.
- *ImageSpace* supports the storage of ontologies and annotations in a relational database.
- Finally, we have developed a simple web-based image retrieval system to search images.

*Organization*. The rest of the paper is organized as follows. Section 2 describes related work. Section 3 gives a brief primer for the DAML+OIL ontology language. Section 4, section 5 and section 6 present how to create and visualize an image ontology, and annotate images based on the created ontology using *ImageSpace*. Section 7 describes our approach to store ontologies and annotations in relational database. Section 8 gives an overview of a prototype image retrieval system. Finally, Section 9 concludes the paper and presents some future work.

## 2. Related Work

Extensive research has been conducted on the processing, searching and retrieval of images [6]. Recently, due to the vision of the Semantic Web [1, 5] and the important role of ontologies, there is an increasing interest in ontology-based approaches to image processing and early results show that the use of ontologies can enhance classification precision [8] and image retrieval performance [7].

Numerous ontology creation tools have been developed. Among them, *Protégé* (http://protege.stanford.edu/), developed at Stanford University, and *OntoEdit* [9] are two well-known representatives. While some of these tools provide partial support to DAML+OIL, *ImageSpace* provides full support of this language, and integrate image ontology creation, image annotation and display in one framework. The tool is built particularly with image support in mind and features a user-friendly interface support for image display and ontology-driven annotation capabilities.

Recently, independently and concurrently, *Protégé* has released five publicly accessible plugins that provide capabilities for ontology visualization: *ezOWL*, *Jambalaya*, *OntoViz*, *OWLViz*, and *TGViz*. *ezOWL* supports graphical ontology building. *ezOWL* and *OntoViz* have *ERWin*-like views of ontology classes (rectangles with names) with their properties and restrictions ("attribute" fields in rectangles). *Jambalaya* [10] provides nested interchangeable views and nicely implements three zooming approaches: geometric, semantic and fisheye zooming. *OWLViz*, and *TGViz* have graph-like views of ontologies. *OWLGraph* shares a lot of features with these tools but it provides a richer set of views and layouts. For more details of the features of *Protégé*, the reader is referred to http://protege.stanford.edu/plugins/domain_visualization.html.

## 3. A Primer on DAML+OIL

DAML+OIL is a semantic markup language for publishing and sharing ontologies on the World Wide Web. It is developed as an extension of XML [14], RDF [15] and RDF Schema (RDF-S) [16] by providing additional constructs along with a formal semantics. DAML+OIL uses 44 constructs (or XML tags) to define ontologies, classes, properties, individuals, data types and their relationships. In the following, we present a brief overview of the major constructs and refer the reader to [2] for more details.

**Classes.** A class defines a group of individuals that share some properties. A class is defined by *daml:Class*, and different classes can be related by *rdfs:subClassOf* into a class hierarchy. Other relationships between classes can be specified by *daml:sameClassAs, daml:disjointWith*, etc. The extension of a class can be specified by *daml:oneOf* with a list of class members or by *daml:intersectionOf* with a list of other classes.

**Properties.** A property states relationships between individuals or from individuals to data values. The former is called *ObjectProperty* and specified by *daml:ObjectProperty*. The later is called *DatatypeProperty* and specified by *daml:DatatypeProperty*. Similarly to classes, different properties can be related by *rdfs:subPropertyOf* into a property hierarchy. The domain and range of a property are specified by *rdfs:domain* and *rdfs:range* respectively. Two properties might be asserted to be equivalent by *daml:samePropertyAs.* In addition, different characteristics of a property can be specified by *daml:TransitiveProperty, daml:UniqueProperty,* etc.

**Property restrictions.** A property restriction is a special kind of class description. It defines an anonymous class, namely the set of class of all individuals that satisfy the restriction. There are two kinds of property restrictions: *value constraints* and *cardinality constraints*. Value constraints restrict the values that a property can take within a particular class, and they are specified by *daml:toClass, daml:hasClass, etc.* Cardinality constraints restrict the number of values that a property can take within a particular class, and they are specified by *daml:minCardinality, daml:maxCardinality, daml:cardinality*, etc.

Recently, DAML+OIL [2] has been revised into OWL, which is a Web ontology language that has become a W3C recommendation [3].

## 4. Creating an Image Ontology

*ImageSpace* provides a user-friendly interface to the user to create image ontologies. Figure 1 shows a snapshot of creating an image ontology *FamilyAlbum*. The four tabs, labeled by *Ontology*, *Class*, *Property* and *Instance,* facilitate the specification of these components and their relationships in a graphical fashion.

As shown in Figure 1, when the *Class* tab is enabled, the left frame displays the class hierarchy, and the right frame shows the relationships of this class with other classes including restriction classes. With this interface, one can easily insert, delete, and update a class. In addition, using the right frame, one can specify the relationships of this class with other classes. At the right-bottom corner of the right frame, is a panel that corresponds to property restrictions, where a user can specify both value constraints and cardinality constraints. Note that those shaded property restrictions are automatically inherited from their parent classes unless they are overridden. Also note that, since a class might have multiple parents, other parent classes are shown in the *SubClassOf* field.

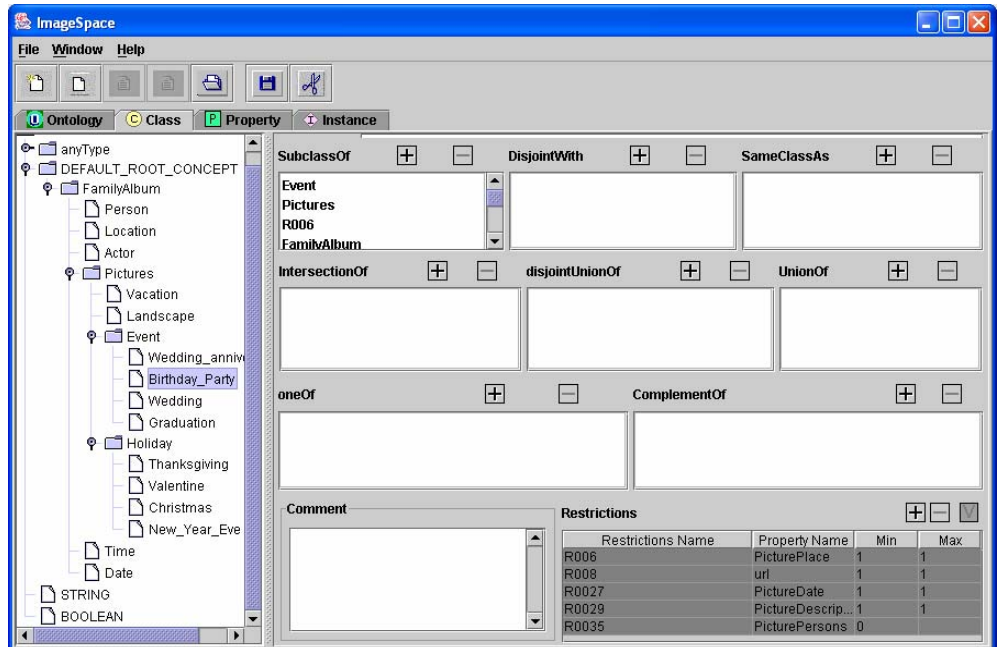

Figure 1. A snapshot of creating image ontology FamilyAlbum
(http://www.cs.wayne.edu/~shiyong/ontology/FamilyAlbum.daml)

When the user enables the *Property* tab, similarly, the left frame shows the property hierarchy, in which parent-child relationship associates the *subPropertyOf* relations between properties. On the right frame, one can specify the type, domain, range of a property. In addition, one can relate a property to other properies in the fields of *InverseOf* and *SamePropertyAs*. Also note that, since a property might have multiple parents, other parent properties are shown in the *SubPropertyOf* field.

Creating restrictions is a part of the definition of a class. It creates an anonymous class. For example, we define a class *Pictures*. Every instance of a class *Pictures* must have a *PicturePlace*. In this case, we define a restriction. Each restriction must have a property called *onProperty*. In other words, that means the restriction is imposed on that property. We can also define a local range using *toClass*, and *hasClass*, and the number for range (*cardinality*, *minCardinality*, *maxCardinality*). The definition of qualification is a part of the restriction. It has *hasClassQ* and the number for range (*cardinalityQ, minCardinalityQ, maxCardinalityQ*). Because restriction is an anonymous class, we represent a restriction with the relation (*subClassOf*, *complementOf*, *unionOf*, *disjoinWith*, *disjointUnionOf*, *sameClassAs*, *intersectionOf*) within the class. For example, when defining a class *Pictures*, *SubClassOf* field contains a restriction on the property *PictureDate*. Its range (*toClass*) is *dateTime* and the number for that range (*cardinality*) is 1. In order to keep the consistency of ontology, we check whether *maxCardinality* is larger than *minCardinality*. If we define the *toClass*, it should not have *hasClass* and qualification; the reverse should agree as well. *Birthday_Party* class (shown on figure 1) has inherited all restrictions from its parent *Pictures*.

The consistency of an ontology is essential and special cares must be taken in order to create a consistent ontology. For example, if Class A is specified as the parent class of Class B, then Class A cannot be in the *complementOf* class list of Class A. *ImageSpace* uses the following four mechanisms to ensure creating only consistent ontologies: (1) *No action.* If an insert, delete or update of a component will violate the consistency of the whole ontology, then the action is canceled with a warning given to the user to indicate the reason of such cancellation. (2) *Cascaded action.* When an offending action occurs, it triggers another or a series of other recovering actions to occur so that the consistency of the ontology is maintained. For example, when a class is deleted, then all references to the class will be deleted as well provided that such cascaded deletion will not cause inconsistency of the ontology. (3) *Using a filter.* To prevent consistency violating action from occurring, a filter is used to restriction the actions that a user can perform. For example, in the *disjointWith* field of a class, a filter is used so that no ancestor classes of this class can be chosen as a class in the *disjointWith* list. (4) *Validation before submission.* This mechanism is used, for example, in the *instance* interface. After an image is annotated, constraints such as cardinality constraints are checked, and if some inconsistencies occur, then the submission is cancelled, with an error message prompted to the user. The submission will not be committed until all constraints are satisfied. A detailed description of all the consistency checks that are performed by *ImageSpace* is beyond the scope of this paper. Interested readers are referred to [4] for details.

## 5. Ontology Visualization

Ontology visualization plays an important role in understanding and maintaining the structure of large knowledge bases. Ontology creation tools usually have many tabs and dialog windows, because of complex relationships and dependencies among classes, properties, and restrictions. As a result, one of the problems that users experience while navigating large ontologies is disorientation.

We have developed a tool for ontology visualization that can work as a stand alone application as well as an *ImageSpace* plugin. It provides simple and user-friendly interface for graphical navigation through ontology.

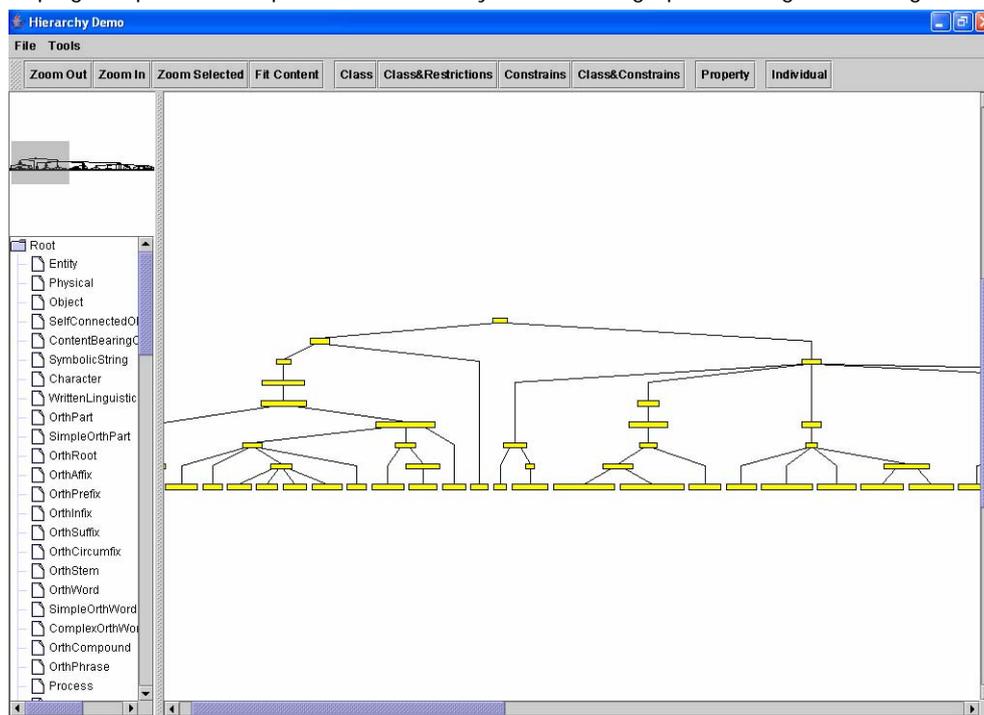

Figure 2. A snapshot of an ontology visualization (class view, hierarchical layout)

Figure 2 shows a snapshot of a sample ontology visualization. The tool main window has a menu, a toolbar, and 3 frames: left upper frame shows preview of a whole ontology graph; right frame shows main view of an ontology; left bottom frame shows a list of classes. A user can use all 3 frames to navigate an ontology.

The visualization plugin supports the following views/hierarchies: class; class and restrictions; constrains; class and constrains; property; and individual. Various concepts (class, property, individual) have different coloring scheme. In addition, a user can experiment with 3 highly customizable layouts: hierarchical, orthogonal, and organic. Figure 2 shows a class view of an ontology displayed with hierarchical layout.

Finally, we provide support of such common features like zooming (in, out, selected content, frame fitting) and manual layout of graphical primitives.

## 6. Annotating an Image

One attractive feature of *ImageSpace* is that, it nicely integrates annotation of images into one framework. The *Instance* tab corresponds to this functionality. Figure 3 displays a snapshot of annotating an image using *ImageSpace*. The left frame shows the class hierarchy and instances (shown by I-icons) associated with the classes to which they belong. The interface on the right frame is ontology-driven. In other words, for different ontologies and different classes, the interface will be generated dynamically based on the properties, cardinality constraints specified for the ontology. For example, for the *FamilyAlbum* ontology, the interface will contain fields *PicturePersons*, *PictureDate*, *PictureDescription* (hidden), etc. While *PictureDate* and *PictureDescription* are *DatatypeProperties*, *PicturePersons* is an *ObjectProperty* that relates an image to a list of actors. Here, an *actor* models a particular snapshot of a person in a particular picture. In the example given, there are two actors. The + button on the right of *PicturePersons* field allows a user to pop up a dialogue window to choose from a list of actors, in which the +/- buttons facilitates the insert/delete of actors in this list. This nested dialogue interface greatly facilitates a user to create instances in an on-the-demand fashion. For example, the insert of an actor might require a person to be inserted first, the nesting order of the dialogue windows ensures that a referenced instance is inserted before a referencing instance is inserted. In our example, the *FamilyAlbum* ontology will enable a user to model that two actors, say *Kathleen-actor1* and *Kevin-actor1*, exist in the picture, that these two actors are for persons *Kathleen* and *Kevin*, and *Kathleen-actor1* hugs *Kevin-actor1* in the picture. In this way, an intelligent semantic search such as "*return all the vacation pictures in which Kathleen hugs Kevin*" can be supported.

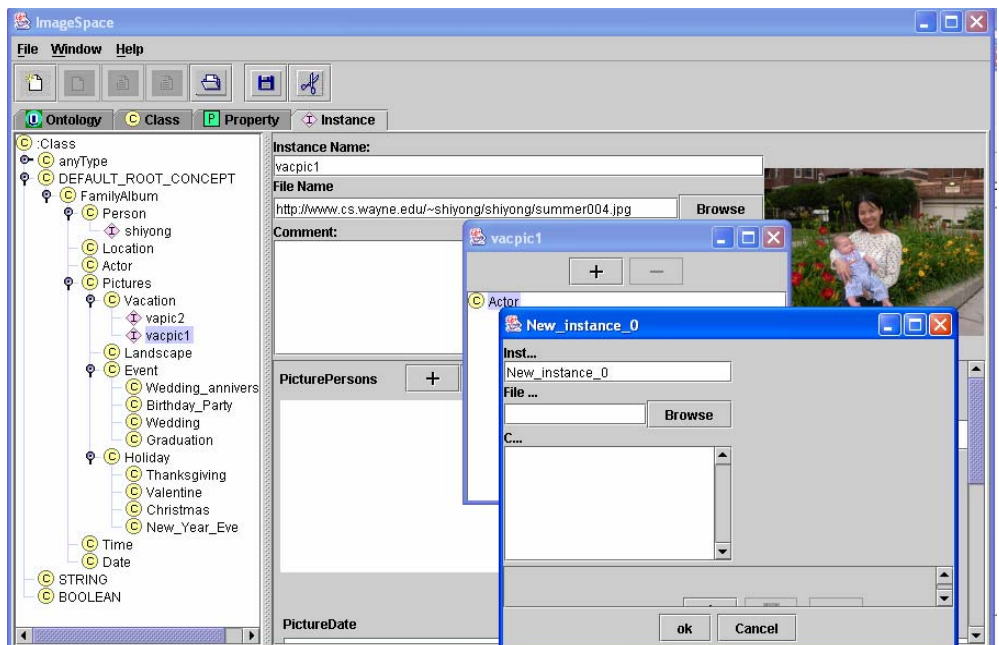

Figure 3. A snapshot of annotating an image

## 7. Storing Ontologies and Annotations in a Relational Database

Both ontologies and annotations are saved in a relational database for the support of ontology-driven search of images. We describe our database design in terms of the following tables that we create where primary keys are underlined:

- Ontology(<u>OntologyID</u>, versionInfo, comment)
- Import(<u>OntologyID, importedOntologyID</u>)
- Class(<u>classID, ontologyID,</u> type, label, comment)
- SubClassOf(<u>classID, parentClassID</u>)
- DisjointWith(<u>classID, otherClassID</u>)
- DisjointUnionOf(<u>classID, otherClassID</u>)
- UnionOf(<u>classID, otherClassID</u>)

- SameClassAs(classID, otherClassID)
- IntersectionOf(classID, otherClassID)
- ComplementOf(classID, otherClassID)
- OneOf(classID, instanceID)
- Property(propertyID, ontologyID, type, comment)
- SubPropertyOf(propertyID, parentPropertyID)
- PropertyDomain(propertyID, classID)
- PropertyRange(propertyID, classID)
- SamePropertyAs(propertyID, otherPropertyID)
- InserseOf(propertyID, classID)
- Restriction(restrictionID, onProp, toClass, minC, maxC, C)
- HasClass(restrictionID, classID)
- HasValue(restrictionID, value)
- HasClassQ(restrictionID, classID, minC, maxC, C)
- Instance(instanceID, classID)
- InstanceRelationship(instanceID, propertyID, value)
- DifferentInvividualFrom(instanceID, otherInstanceID)
- SameIndividualAs(instanceID, otherInstanceID)

As an example, consider an image where Kathleen smiles and hugs Kevin, and Kevin cries. An appropriate annotation can be stored in relational tables *Instance* and *InstanceRelationship* which are shown in table 1 and table 2 correspondingly. In practice, for efficiency concerns, we split *InstanceRelationship* table to set of tables with names that correspond to *propertyID* attribute value and with attributes *subject* (corresponds to *instanceID*) and *value*. Thus, the final schema for our example will contain the following tables (instead of *InstanceRelationship*):

- hasActor (subject, value)
- hugs (subject, value)
- hasAction (subject, value)
- hasName (subject, value)
- isSnapshotOf (subject, value)

Table 1. Relational table Instance

| instanceID | classID |
|---|---|
| Kathleen | Person |
| Kevin | Person |
| http://www.cs.wayne.edu/example.jpg | Vacation |
| Kathleen-actor1 | Actor |
| Kevin-actor1 | Actor |

Table 2. Relational table InstanceRelationship

| instanceID | propertyID | value |
|---|---|---|
| http://www.cs.wayne.edu/example.jpg | hasActor | Kathleen-actor1 |
| http://www.cs.wayne.edu/example.jpg | hasActor | Kevin-actor1 |
| Kathleen-actor1 | hugs | Kevin-actor1 |
| Kathleen-actor1 | hasAction | smiles |
| Kevin-actor1 | hasAction | cries |
| Kathleen | hasName | Kathleen |
| Kevin | hasName | Kevin |
| Kathleen-actor1 | isSnapshotOf | Kathleen |
| Kevin-actor1 | isSnapshotOf | Kevin |

## 8. Ontology-based Image Retrieval

Based on this database schema presented in the previous section, we have developed a simple web-based image retrieval system to search images. The system provides an interface to allow the user to navigate to images under different categories. In addition, a user can specify a list of "triples" as the search criterion to retrieve images. For example, one can specify a search criterion such as return all the images under the "vacation" category such that:
- Kathleen hugs Kevin, and
- Kathleen smiles, and
- Kevin cries.

The following datalog-style query will retrieve the needed photos where variables are prefixed by a '$':

Answer ($instanceID) :-
    instanceOf ($instanceID, Vacation),
    hasActor ($instanceID, $A1),
    hasActor ($instanceID, $A2),
    isSnapshotOf ($A1, $P1),
    isSnapshotOf ($A2, $P2),
    hasName ($P1, "Kathleen"),
    hasName ($P1, "Kevin"),
    hugs ($A1, $A2),
    hasAction ($A1, smiles),
    hasAction ($A2, cries).

Finally, query is translated to the following sequence of SQL statements:
- Select all actors for "Kathleen" and store them into *KathleenActor*.
  SELECT isSnapshotOf.subject
  FROM isSnapshotOf, hasName
  WHERE isSnapshotOf.value = hasName.subject AND hasName.value = 'Kathleen'
- Select all actors for "Kevin" and store them into *KevinActor*.
  SELECT isSnapshotOf.subject
  FROM isSnapshotOf, hasName
  WHERE isSnapshotOf.value = hasName.subject AND hasName.value = 'Kevin'
- Select all "smiling" actors for "Kathleen" and store them into *SmilingKathleenActor*.
  SELECT hasAction.subject
  FROM KathleenActor, hasAction
  WHERE KathleenActor.subject = hasAction.subject AND hasAction.value = 'smiles'
- Select all "crying" actors for "Kevin" and store them into *CryingKevinActor*.
  SELECT hasAction.subject
  FROM KevinActor, hasAction
  WHERE KevinActor.subject = hasAction.subject AND hasAction.value = 'cries'
- Retrieve all images that satisfy all specified conditions.
  SELECT H1.subject
  FROM hasActor H1, hasActor H2, Hugs
      SmilingKathleenActor, CryingKevinActor
  WHERE H1.subject = H2.subject AND
      H1.value = SmilingKathleenActor.subject AND
      H2.value = CryingKevinActor.subject AND
      Hugs.subject = SmilingKathleenActor.subject
      AND Hugs.value = CryingKevinActor.subject

All and only the images that satisfy this criterion will be returned (in our case, http://www.cs.wayne.edu/example.jpg). The reader is referred to [4] for more details about the ontology-driven image retrieval system.

## 9. Conclusions and Future Work

We have developed *ImageSpace*, an image ontology creation, visualization and annotation tool that fully supports the standard DAML+OIL ontology language and enables the storage of ontologies and annotations in a relational database. Future work includes:
- The development of *MultimediaSpace* that will not only support the annotation of images, but also other multimedia resources such as videos, audios, etc.
- Future version of *MultimediaSpace* will also support OWL, the successor of DAML+OIL.

- The development of graphical ontology building features to support by *MultimediaSpace* and visualization plug-in.
- Better optimization of SQL queries that are generated by image retrieval system.

## Authors' Information

**Shiyong Lu** – e-mail: shiyong@cs.wayne.edu

**Rong Huang** – e-mail: f10272@cs.wayne.edu

**Artem Chebotko** – e-mail: artem@cs.wayne.edu

**Yu Deng** – e-mail: yudeng@cs.wayne.edu

**Farshad Fotouhi** – e-mail: fotouhi@cs.wayne.edu

Department of Computer Science, Wayne State University, Detroit, MI 48202, USA